\documentclass[14pt]{article}

\usepackage{arxiv}
\usepackage{amsmath,amssymb}

\usepackage[utf8]{inputenc} % allow utf-8 input
\usepackage[T1]{fontenc}    % use 8-bit T1 fonts
\usepackage{hyperref}       % hyperlinks
\usepackage{url}            % simple URL typesetting
\usepackage{booktabs}       % professional-quality tables
\usepackage{amsfonts}       % blackboard math symbols
\usepackage{nicefrac}       % compact symbols for 1/2, etc.
\usepackage{microtype}      % microtypography
\usepackage{lipsum}
\usepackage{graphicx}
\usepackage{multirow}
\usepackage{subcaption}
\usepackage{natbib}
\graphicspath{ {./images/} }

\title{Two-Stage Estimation of Population Abundance with Robust Inference under Interacting Survey Protocols}

\author{
 Yusaku Ohkubo \\
  Okayama University\\
  Okayama, Japan \\
  \texttt{y-ohkubo@okayama-u.ac.jp} \\
       \And
 Tatsuki Shimamoto\\
  Nippon Veterinary and Life Science University\\
  Tokyo, Japan \\
  \texttt{pteromys.volans.momonga@gmail.com}\\
           \And
 Yuya Eguchi \\
  Azabu University\\
  Kanagawa, Japan \\
  \texttt{yuya.sciurus.09@gmail.com} \\
     \And
 Hirotaka Katahira \\
  Azabu University\\
  Kanagawa, Japan \\
  \texttt{katahira@azabu-u.ac.jp} \\
}

\begin{document}
\maketitle
\begin{abstract}
Estimating population abundance from field surveys is often complicated by interference between multiple survey protocols. In this paper, we propose a two-stage estimation framework for abundance models in which detection processes interact, leading to both missed detections and sample loss caused by survey procedures. Our approach separates the calibration of sample loss from the estimation of detection probability, thereby avoiding the feedback and weak identifiability that can arise in fully joint hierarchical models. We further derive a sandwich-type robust variance estimator that propagates first-stage uncertainty into the second stage and remains valid under certain forms of model misspecification. Simulation studies demonstrated that the proposed method provides more reliable uncertainty quantification than a Bayesian hierarchical joint model, which tends to underestimate uncertainty even under correct specification. We illustrate the practical utility of the method using ectoparasite abundance data from the invasive Pallas’s squirrel \textit{Callosciurus erythraeus}.
\end{abstract}

% keywords can be removed

\section*{Data Availability}
The data that support the findings of this study are available from the corresponding author upon reasonable request.

\section*{Acknowledgement}
This research was partially supported by JSPS KAKENHI (Grant Numbers 21K15170 and 24K15120). The core ideas of this study were presented at the Fourth Seminar of the School of Statistical Thinking, the Institute of Statistical Mathematics. 

\section*{Conflicts of Interest}
The authors declare no conflicts of interest.

\fontsize{12}{13.5}\selectfont
\clearpage
\section{Introduction}
Estimating the abundance of a population is fundamental in ecology, yet it is often challenging in practice \citep{KerySchaub2011}. Because inspecting an entire population without error is unrealistic, many sampling schemes have been proposed to collect abundance data in the field \citep{kellner2014accounting}. Each method inevitably incurs costs, and the quality of the resulting data varies. Typically, a more accurate method requires greater expenditure of human effort, time, or other resources than a less accurate one. For example, aerial surveys, which count target animals from an airplane, are known to provide a relatively accurate estimate of abundance though expensive, while simple transect methods are cheaper and applicable to wider areas with a risk of intrinsic bias \citep{Hone2008}. Applying both methods to the same small area allows us to assess the efficiency and accuracy of the low-cost, lower-quality method—that is, how reliable it is compared with the high-cost method—while the low-cost method can be applied to a larger area and its data subsequently calibrated. This combined approach offers a practical strategy to optimize resource use while maintaining acceptable data quality.

However, the situation becomes more complicated when the combination of a high-cost, high-quality method and a low-cost, low-quality method introduces an additional source of bias. For example, aerial surveys using UAVs can generate noise that stresses animals \citep{MuleroPazmany2017}, while artificial lighting used for nighttime counts may attract or repel focal species \citep{RyerOlla1999}, among other potential artifacts \citep{Boulanger}. Such effects can induce heterogeneity in detection efficiency or systematically alter the observation process of subsequent surveys, violating standard assumptions of independent and non-interfering sampling. When one protocol affects the data-generating mechanism of another, naïve calibration strategies may lead to biased estimation and misleading uncertainty quantification. Such interactions between sampling methods highlight the necessity of statistical approaches that can explicitly account for a combination of heterogeneous survey protocols.

A natural strategy for handling heterogeneity in detection probability is the use of N-mixture models, also known as Poisson–Binomial models \citep{Royle2004}. N-mixture models are hierarchical frameworks in which a latent Poisson variable represents the true abundance of a population, while imperfect detection is modeled through an observation process. These models have been extended in various ways and widely applied in empirical ecological studies\citep{KerySchaub2011}. Despite their popularity, however, previous work has shown that N-mixture models can suffer from substantial bias under sparse data and are sensitive to model misspecification, particularly when the outcome data follows a negative binomial distribution \citep{Barker2018, Knape2018, Link2018, Veech2016}. Although increasing the number of replicate surveys per site may alleviate these issues, doing so often limits practical applicability, particularly when ecologists aim to combine a high-cost, high-accuracy survey method with a low-cost, lower-accuracy alternative. In such settings, additional replications may be infeasible, and more robust estimation strategies are required. 

\subsection{Motivating Example}
Abundance of parasite is a key component for understanding parasite ecology and accurately assessing pathogen transmission. In particular, invasive species have attracted considerable attention because they can co-introduce non-native pathogens and their vectors (\citep{Zhang2022}. Information on parasite presence and abundance is crucial not only for detecting parasite invasions into native ecosystems \citep{Goedknegt2016}\citep{Poulin2017}, but also for assessing the risk of zoonotic pathogen transmission to public health from the One Health perspective \citep{Chinchio2020}. 

For instance, population and distribution of Pallas’s squirrels \textit {Callosciurus erythraeus} have been expanding in some regions in Japan. Pallas’s squirrels are classified as arboreal squirrels and their typical habitat is mostly forest; they, however, inhabit even in urban green spaces both in native and introduced habitats \citep{Seki2022}\citep{Thaweepworadej2023}. Therefore, their occupancy in residential areas and urban environments may increase opportunities for contact among squirrels, native species, companion animals, and humans, potentially facilitating the spread of parasites and pathogen transmission. Some of the ectoparasite, which inhabit the outer surface of their terrestrial mammal hosts, detected from Pallas’s squirrels are non-native lice and native ticks, mites, and fleas \citep{Katahira2022}. These species are known vectors of zoonotic pathogens and also have a potential risk of pathogen pollution. Obtaining accurate estimate of  parasite abundance in Pallas’s squirrels is essential to providing risk assessment of these pathogen transmissions.

To estimate the abundance of ectoparasites, two common sampling methods—the shaving method and the combing method—have been applied to collect individual parasites. The shaving method uses a razor to remove the host’s fur in order to collect parasites attached to it. In contrast, the combing method involves manually combing the host’s fur for a fixed period of time, typically 5–10 minutes, to collect parasites. Because whole-body shaving is highly labor- and time-intensive (e.g., shaving the entire body of a squirrel may take a full day, followed by several additional days to examine the removed fur for ectoparasites), the complementary use of shaving and combing may be worth considering when estimating ectoparasite abundance.

Combing sampling method, however, poses several difficulties in practice. It tends to underestimate the actual abundance of a population because some individuals present on the host remain undetected \citep{Hopkins1949}. One approach to correct this bias is to conduct supplementary “calibration” experiments. In such experiments, the combing method is followed by the shaving method, which is assumed to detect nearly all remaining parasites. By comparing parasite counts from the combing and shaving steps within the calibration data, researchers can estimate the detection probability of the combing method, estimated as the number of parasites found by combing divided by the total number found after shaving. The observed counts from combing method can then be adjusted by multiplying them by the inverse of this estimated detection probability. Unfortunately, these calibrated data are still insufficient to obtain accurate estimates of parasite abundance. One important limitation is that the combing method may physically damage or remove parasite individuals, resulting in further underestimation. Indeed, our real-world data analysis reveals that the number of parasites detected by the shaving method after combing was consistently smaller than the number detected by shaving alone, even when host individuals were randomly assigned to each groups and relevant covariates (e.g., host sex and body size) were controlled for. This indicates that parasite individuals are lost during the combing method and that the detection probability of the combing method is overestimated. We need to account for two sources of bias to obtain accurate and reliable estimates of parasite abundance: (1) missed detections—parasites that remain on the host after combing—and (2) sample destruction—parasites damaged or removed during combing but not detected by subsequent methods. Although our motivating example arises from these issues, the statistical problem we address (i.e. interaction of survey protocols with heterogeneous detection and loss mechanisms) is common across many applied fields, including epidemiology and environmental monitoring.

\subsection{Outline of the Paper}
In this paper, we propose a two-stage estimation framework for interacting detection processes to address these issues by extending the N-mixture model. To accommodate the structure of our data, we develop a model that explicitly represents heterogeneity in the effective detection process arising from (i) missed detections and (ii) sample loss caused by procedural damage. Building on this model, we propose a two-stage estimator. In the first stage, calibration data are analyzed to estimate the number of parasite individuals that were likely missed due to sample loss, yielding a counterfactual prediction of the undamaged abundance. In the second stage, the detection probability of the combing method is estimated conditional on this counterfactual abundance. Estimation of unknown parameters can be implemented within the usual maximum likelihood estimation framework of Poisson generalized linear models, owing to a marginal representation of the N-mixture model. We further derive a robust variance estimator that formally accounts for first-stage uncertainty and remains valid under certain forms of model misspecification. Our variance estimator can be viewed as an extension of the classical sandwich estimator to two-stage estimating equations with generated regressors. The performance of the proposed estimator is evaluated through simulation studies. Finally, we apply the method to real data on ectoparasite counts from the invasive Pallas’s squirrel \textit {C. erythraeus}.

\section{Proposed Method}

We developed a novel statistical method for estimating  abundance in the presence of two distinct sources of bias. Total $N$ populations are assigned to three groups, with sizes denoted by $N_{\mathrm{ctrl}}$, $N_{\mathrm{trt}}$, and $N_{\mathrm{test}}$. The control group is examined using a high-accuracy method (e.g., shaving), and the observed count is recorded as $y_{\mathrm{ctrl}}$. In the treatment group, a low-accuracy method (e.g., combing) is applied first, yielding $y_{\mathrm{low}}$, followed by the high-accuracy method, yielding $y_{\mathrm{high}}$. Each low-accuracy survey consists of $Z_i\in\{0,1,2,\dots\}$ repeated sessions. The test group is examined using only the low-accuracy method, yielding $y_{\mathrm{test}}$. Optionally, covariates $X_i$ (e.g., environmental conditions) may be recorded for each populations.

A naïve estimator of the per-session detection probability of the low-accuracy method, $p_{\mathrm{low}}$, would model the treated group via
\[
y_{\mathrm{low},j} \sim \mathrm{Binomial}\!\left(y_{\mathrm{high},j},\, p_{\mathrm{low}}\right),
\]
implicitly assuming that the high-accuracy method detects nearly all individuals. Maximum likelihood would yield
\[
\widehat{p}_{\mathrm{low}} = \frac{y_{\mathrm{low}}}{y_{\mathrm{high}}}.
\]
However, this estimator is biased upward because the low-accuracy method may physically remove or damage parasites prior to the high-accuracy measurement. Thus, $y_{\mathrm{high}}$ does not represent the full remaining population, leading to systematic overestimation of $p_{\mathrm{low}}$ and underestimation of abundance when applied to $y_{\mathrm{test}}$.

To address this issue, we decompose the inference into two stages:  
(1) predicting the number of individuals lost due to low-accuracy sampling, and  
(2) estimating $p_{\mathrm{low}}$ conditional on this prediction.

We model the measurement outcomes as
\begin{align}
y_{\mathrm{ctrl},i} &\sim \mathrm{Binomial}\!\left(y_{\mathrm{TRUE},i},\, 1\right), \qquad i=1,\dots,N_{\mathrm{ctrl}},\\[3pt]
y_{\mathrm{high},j} &\sim \mathrm{Binomial}\!\left(y_{\mathrm{TRUE},j},\, p_{\mathrm{intact}}^{Z_j}\right), \qquad j=1,\dots,N_{\mathrm{trt}}, \\[3pt]
y_{\mathrm{low},j} &\sim \mathrm{Binomial}\!\left(y_{\mathrm{TRUE},j},\, 1-(1-p_{\mathrm{low}})^{Z_j}\right), \qquad j=1,\dots,N_{\mathrm{trt}},
\end{align}
where $p_{\mathrm{intact}}$ is the per-session survival probability under the low-accuracy method, $p_{\mathrm{low}}$ is the per-session detection probability of the low-accuracy method, and $Z_j$ is the number of low-accuracy sessions applied to individual $j$.

The latent abundance is modeled as
\begin{align}
y_{\mathrm{TRUE},i} \sim \mathrm{Poisson}(\lambda_i),
\qquad
\log \lambda_i = X_i^\top\beta.
\end{align}

where $\beta$ represents the regression coefficients. This hierarchical model is a special case of the N-mixture model \citep{Royle2004}, except that the control group has perfect detection probability.

We consider two approaches to estimate the unknown parameters. The first approach is a Bayesian hierarchical joint model that represents Eqs. (2.1)–(2.4) and estimates all unknown parameters simultaneously. A standard Markov Chain Monte Carlo (MCMC) algorithm can be applied in this setting. The second, and our proposed, approach is a two-stage estimator, which divides the model into two parts and estimates the parameters step by step.

At the first-stage, noting a marginal distribution of $y_{\mathrm{high},j}$, we model the control and high-accuracy counts as
\begin{align}
y_{\mathrm{ctrl},i}  &\sim \mathrm{Poisson}(\lambda_i),\\
y_{\mathrm{high},j} &\sim \mathrm{Poisson}(\lambda_j\, p_{\mathrm{intact}}^{Z_j}),
\end{align}
and estimate $(\beta,p_{\mathrm{intact}})$ by maximum likelihood. Intuitively, repeated low-accuracy sessions reduce the expected high-accuracy count multiplicatively, and $p_{\mathrm{intact}}^{Z_j}$ captures this loss.  
A counterfactual abundance prediction (“no loss”) is obtained as
\begin{align}
\widehat{\lambda}_{\mathrm{pred},j} = \exp(X_j^\top \widehat\beta),
\end{align}
corresponding to $p_{\mathrm{intact}}=1$. This counterfactual prediction serves to reconstruct the “true” parasite abundance that would have been observed in the absence of combing-induced loss, and to quantify the expected number of individuals lost by this damage, by comparing the observed abundance in the treatment group with its counterfactual one.

At the second-stage, conditional on this predicted abundances, we model
\begin{align}
y_{\mathrm{low},j}
\sim
\mathrm{Poisson}\!\left(
\widehat{\lambda}_{\mathrm{pred},j}\,
\big[1-(1-p_{\mathrm{low}})^{Z_j}\big]
\right),
\end{align}
yielding the estimator $\widehat{p}_{\mathrm{low}}$ by MLE.

Stage-1 estimation reduces to a standard Poisson GLM of the form
\begin{align}
\log \widetilde{\lambda}_j
= X_j^\top \beta + Z_j\log p_{\mathrm{intact}}.
\end{align}
So consistency and asymptotic normality follow from classical GLM theory.

In contrast, Stage-2 uses predicted intensities $\widehat{\lambda}_{\mathrm{pred}}$, whose uncertainty must be propagated. Ignoring this produces underestimate of standard errors. We therefore derive a robust, sandwich-type variance estimator for $\widehat{p}_{\mathrm{low}}$ that accounts for first-stage uncertainty. See the Appendix for details.

\section{Simulation Study}
We conducted numerical simulations to investigate the statistical properties of the proposed method. Data were generated according to the model described in Section 2 (equations 2.1–2.5). The total sample size was $N\in{\{150,300}\}$, with $N_\text{trt}=2N_\text{ctr}$. Each covariate vectors $x_i=(x_{i1},\ x_{i2},...x_{i5})^T$ had the first component fixed at $1$ (intercept) and the remaining four components independently drawn from $x_i\ \sim \ \mathrm{Normal}(0,\ 1)$. The true regression coefficients were fixed at ${\beta}=(1,\ 0.5,\ 0.5,\ -0.5,\ -0.5)^T$. The variable $z_i$ was set to $0$ for the first one-third of the units, $1$ for the second one-third, and $2$ for the final one-third. 

We applied two approaches to the simulated data: (1) the proposed two-stage method and (2) the posterior mean value from the Bayesian hierarchical joint model, which correctly captures the true data-generating process. In the second stage of the proposed method, we computed two types of standard errors (SEs): the usual SE from the inverse observed Fisher information and the Humberger-type robust SE, which we derived. For the hierarchical joint model, we specified the following priors:

\begin{align*}
    {\beta}\ \sim \ \mathrm{Normal}(0,\ 10^6)\\
    p_\text{intact}\ \sim \ \mathrm{Uniform}(0,\ 1)\\
    {\ p}_\text{low}\ \sim \ \mathrm{Uniform}(0,\ 1)
\end{align*}

where $\mathrm{Uniform}\left(\alpha,\beta\right)$ denotes the uniform distribution on $[\alpha,\ \ \beta]$.

This process was repeated 1,000 times, and we evaluated the Frequentist properties, including empirical bias (i.e., the average difference from the true value), root mean squared error (RMSE), and coverage rate (i.e., the proportion of times the 95 percent interval for a parameter contained the true value). We tested $3\ \times4$ combinations of the true parameter values $(p_\text{intact}\in{0.7,\ 0.8,\ 0.9},\ p_\text{low}\in{0.05,\ 0.1,\ 0.15,\ 0.2})$ and two different sample sizes. All analyses were conducted in \texttt{R} (version 4.3) using the packages \texttt{R2jags}.

\begin{table}[htbp]
\centering
\caption{Comparisons of bias and RMSE between the two types of estimators for different sample sizes ($N\in \left\{150,\ 300\right\})$ and for 12 different scenarios of the true values. Each cell reports the average of the point estimates, with the corresponding RMSE shown in parentheses. The Bayesian hierarchical joint model tends to overestimate the impact of the treatment variable, while the proposed methods achieved unbiased estimates. Moreover, although not uniformly across all scenarios, the proposed methods achieve smaller RMSE than the joint model in the majority of settings.}
\begin{tabular}{c c c c c c c}
\hline
$N$ & \multicolumn{2}{c}{TRUE VALUE} & \multicolumn{2}{c}{BAYES} & \multicolumn{2}{c}{PROPOSED} \\
\cline{2-7}
& $p_{\text{intact}}$ & $p_{\text{low}}$ 
& $p_{\text{intact}}$ & $p_{\text{low}}$ 
& $p_{\text{intact}}$ & $p_{\text{low}}$ \\
\hline
\multirow{12}{*}{150}
& 0.7 & 0.05 & 0.713  {\scriptsize(0.047)} & 0.053  {\scriptsize(0.010)} & 0.700  {\scriptsize(0.043)} & 0.050  {\scriptsize(0.010)}\\
&     & 0.10 & 0.723 \ {\scriptsize(0.049)} & 0.106 \ {\scriptsize(0.017)} & 0.700 \ {\scriptsize(0.041)} & 0.100 \ {\scriptsize(0.015)}\\
&     & 0.15 & 0.735 \ {\scriptsize(0.058)} & 0.162 \ {\scriptsize(0.025)} & 0.700 \ {\scriptsize(0.041)} & 0.151 \ {\scriptsize(0.020)}\\
&     & 0.20 & 0.756 \ {\scriptsize(0.074)} & 0.222 \ {\scriptsize(0.038)} & 0.703 \ {\scriptsize(0.040)} & 0.202 \ {\scriptsize(0.024)}\\
\cline{2-7}

& 0.8 & 0.05 & 0.816 \ {\scriptsize(0.048)} & 0.054 \ {\scriptsize(0.011)} & 0.802 \ {\scriptsize(0.045)} & 0.050 \ {\scriptsize(0.010)}\\
&     & 0.10 & 0.824 \ {\scriptsize(0.051)} & 0.106 \ {\scriptsize(0.017)} & 0.803 \ {\scriptsize(0.046)} & 0.101 \ {\scriptsize(0.015)}\\
&     & 0.15 & 0.831 \ {\scriptsize(0.054)} & 0.161 \ {\scriptsize(0.023)} & 0.800 \ {\scriptsize(0.044)} & 0.151 \ {\scriptsize(0.019)}\\
&     & 0.20 & 0.850 \ {\scriptsize(0.068)} & 0.221 \ {\scriptsize(0.035)} & 0.803 \ {\scriptsize(0.045)} & 0.202 \ {\scriptsize(0.024)}\\
\cline{2-7}
& 0.9 & 0.05 & 0.908 \ {\scriptsize(0.038)} & 0.052 \ {\scriptsize(0.010)} & 0.902 \ {\scriptsize(0.047)} & 0.050 \ {\scriptsize(0.009)}\\
&     & 0.10 & 0.913 \ {\scriptsize(0.039)} & 0.104 \ {\scriptsize(0.015)} & 0.902 \ {\scriptsize(0.048)} & 0.101 \ {\scriptsize(0.015)}\\
&     & 0.15 & 0.917 \ {\scriptsize(0.040)} & 0.156 \ {\scriptsize(0.019)} & 0.899 \ {\scriptsize(0.048)} & 0.150 \ {\scriptsize(0.019)}\\
&     & 0.20 & 0.925 \ {\scriptsize(0.041)} & 0.211 \ {\scriptsize(0.025)} & 0.899 \ {\scriptsize(0.047)} & 0.199 \ {\scriptsize(0.024)}\\
\hline
\multirow{12}{*}{300}
& 0.7 & 0.05 & 0.709 \ {\scriptsize(0.032)} & 0.052 \ {\scriptsize(0.007)} & 0.699 \ {\scriptsize(0.030)} & 0.050 \ {\scriptsize(0.007)}\\
&     & 0.10 & 0.718 \ {\scriptsize(0.036)} & 0.104 \ {\scriptsize(0.012)} & 0.700 \ {\scriptsize(0.029)} & 0.100 \ {\scriptsize(0.010)}\\
&     & 0.15 & 0.731 \ {\scriptsize(0.045)} & 0.158 \ {\scriptsize(0.018)} & 0.701 \ {\scriptsize(0.030)} & 0.150 \ {\scriptsize(0.014)}\\
&     & 0.20 & 0.747 \ {\scriptsize(0.058)} & 0.216 \ {\scriptsize(0.026)} & 0.700 \ {\scriptsize(0.029)} & 0.200 \ {\scriptsize(0.017)}\\
\cline{2-7}
& 0.8 & 0.05 & 0.811 \ {\scriptsize(0.033)} & 0.052 \ {\scriptsize(0.007)} & 0.801 \ {\scriptsize(0.030)} & 0.050 \ {\scriptsize(0.007)}\\
&     & 0.10 & 0.818 \ {\scriptsize(0.038)} & 0.104 \ {\scriptsize(0.012)} & 0.800 \ {\scriptsize(0.032)} & 0.100 \ {\scriptsize(0.010)}\\
&     & 0.15 & 0.828 \ {\scriptsize(0.043)} & 0.159 \ {\scriptsize(0.018)} & 0.800 \ {\scriptsize(0.031)} & 0.150 \ {\scriptsize(0.014)}\\
&     & 0.20 & 0.843 \ {\scriptsize(0.054)} & 0.217 \ {\scriptsize(0.026)} & 0.801 \ {\scriptsize(0.031)} & 0.201 \ {\scriptsize(0.017)}\\
\cline{2-7}
& 0.9 & 0.05 & 0.909 \ {\scriptsize(0.032)} & 0.052 \ {\scriptsize(0.007)} & 0.900 \ {\scriptsize(0.032)} & 0.050 \ {\scriptsize(0.007)}\\
&     & 0.10 & 0.917 \ {\scriptsize(0.034)} & 0.104 \ {\scriptsize(0.011)} & 0.900 \ {\scriptsize(0.033)} & 0.100 \ {\scriptsize(0.010)}\\
&     & 0.15 & 0.923 \ {\scriptsize(0.037)} & 0.157 \ {\scriptsize(0.015)} & 0.900 \ {\scriptsize(0.033)} & 0.150 \ {\scriptsize(0.014)}\\
&     & 0.20 & 0.932 \ {\scriptsize(0.041)} & 0.213 \ {\scriptsize(0.021)} & 0.900 \ {\scriptsize(0.032)} & 0.201 \ {\scriptsize(0.016)}\\
\hline
\end{tabular}
\end{table}

Table 1 shows the results of the point estimates. For the proposed method, the mean of the 1,000 estimates were almost identical to the true value, indicating that the estimator is an unbiased. In contrast, the mean estimates obtained from Bayesian joint hierarchical model were systematically biased. These results suggest that Bayesian joint hierarchical model yield misleading estimate, even when the model captures the true data-generating process. The existence of bias is consistent across the scenarios and tends to be larger as the sample size increases.

Table 2 shows the coverage rates of the $95 \%$  intervals. For the proposed methods, the empirical coverage rates were close to the nominal $95 \%$ across scenarios, indicating that the derived robust SE provides a reliable measure of uncertainty. In contrast, Bayesian joint hierarchical model and naïve SE showed lower coverage rates, suggesting that they are inadequate from a Frequentist perspective. Even though coverage rates of joint hierarchical model are better when the true${\ p}_\text{low}$ is small, it can be as low as 50 percent when true${\ p}_\text{low}$ is relatively large. Again, increasing sample size does not improve the performance of Bayesian joint hierarchical model. 

The poor performance of the joint model can be explained by feedback from the observation model of the $y_{\text{low}}$ counts into the estimation of $p_{\text{intact}}$. While our goal is to estimate the detection probability of the combing method conditional on the correction for sample loss, the joint likelihood allows $y_{\text{low}}$ to inform both processes simultaneously, inducing strong posterior dependence and weak identifiability. This type of feedback is conceptually related to concerns raised in the propensity score literature, where joint modeling of the treatment assignment mechanism and the outcome model may induce unintended information flow from the outcome back to the propensity score (\citep{Zigler2013}). In contrast, the two-stage approach enforces a unidirectional flow of information from the calibration experiment to the detection model, thereby avoiding the feedback that arises in the joint model.

We further evaluated the impact of model misspecification on both the proposed two-stage method and the Bayesian hierarchical joint model, where both estimation procedures continued to assume a Poisson likelihood. First, we considered misspecification arising from over-dispersion in the latent intensity of the count process, a common feature of ecological abundance data. The mean intensity was specified as $\lambda_i = \exp(Z_i^\top \beta + r_i), \ r_i \sim N(0,\sigma^2)$, with $\sigma = 0.3$. Second, we considered a setting in which the outcome data followed a negative binomial distribution, as $y_i \sim \mathrm{NegBin}\bigl(\mu_i=\exp(Z_i^\top\beta),\; k\bigr), k=0.3$. In these misspecified scenarios, we intentionally applied robust variance estimation at both stages of the proposed procedure to assess robustness to violations of distributional assumptions, rather than efficiency under correct specification. 

Table 3-4 reports the coverage probabilities of the $95\%$ intervals under this misspecified setting. The results indicate that coverage deteriorates for the Bayesian hierarchical joint model. In contrast, the proposed two-stage estimator maintains coverage at or above the nominal level, albeit with slightly conservative intervals in some cases. These findings suggest that the proposed approach is less sensitive to unmodeled heterogeneity and provides more reliable uncertainty quantification under variance misspecification.
    
\begin{table}[htbp]
\centering
\caption{Comparisons of coverage rate between the three types of estimators. The proportion of a $95\% $ interval that covers the true value is reported. The Bayesian hierarchical joint model tends to be too optimistic about the uncertainty of the interval estimate, while the naïve SE of $p_\text{intact}$ and robust SE of ${\ p}_\text{low}$ are closer to the nominal rate. For $p_{\text{intact}}$, the variances of the first-stage estimator are identical under both the naïve and proposed approaches, and therefore no separate entry is shown for the proposed method.}
\begin{tabular}{c c c c c c c c c}
\hline
$N$ & \multicolumn{2}{c}{TRUE} & \multicolumn{2}{c}{BAYES} & \multicolumn{2}{c}{NAÏVE SE} & \multicolumn{2}{c}{PROPOSED} \\
\cline{2-9}
& $p_{\text{intact}}$ & $p_{\text{low}}$ 
& $p_{\text{intact}}$ & $p_{\text{low}}$ 
& $p_{\text{intact}}$ & $p_{\text{low}}$ 
& $p_{\text{intact}}$ & $p_{\text{low}}$ \\
\hline
\multirow{12}{*}{150}
& 0.7 & 0.05 & 0.941 & 0.947 & 0.950 & 0.937 &--  & 0.947 \\
&     & 0.10 & 0.923 & 0.937 & 0.956 & 0.924 &--  & 0.945 \\
&     & 0.15 & 0.864 & 0.913 & 0.963 & 0.901 &--  & 0.951 \\
&     & 0.20 & 0.766 & 0.862 & 0.961 & 0.893 &--  & 0.947 \\
\cline{2-9}
& 0.8 & 0.05 & 0.932 & 0.932 & 0.954 & 0.922 &--  & 0.932 \\
&     & 0.10 & 0.905 & 0.935 & 0.957 & 0.921 &--  & 0.947 \\
&     & 0.15 & 0.896 & 0.918 & 0.951 & 0.924 &--  & 0.956 \\
&     & 0.20 & 0.793 & 0.865 & 0.963 & 0.895 &--  & 0.958 \\
\cline{2-9}
& 0.9 & 0.05 & 0.955 & 0.950 & 0.939 & 0.940 &--  & 0.947 \\
&     & 0.10 & 0.942 & 0.934 & 0.938 & 0.902 &--  & 0.935 \\
&     & 0.15 & 0.928 & 0.936 & 0.922 & 0.908 &--  & 0.947 \\
&     & 0.20 & 0.895 & 0.925 & 0.931 & 0.889 &--  & 0.947 \\
\hline
\multirow{12}{*}{300}
& 0.7 & 0.05 & 0.934 & 0.941 & 0.942 & 0.943 &--  & 0.951 \\
&     & 0.10 & 0.909 & 0.937 & 0.954 & 0.927 &--  & 0.953 \\
&     & 0.15 & 0.813 & 0.901 & 0.953 & 0.892 &--  & 0.941 \\
&     & 0.20 & 0.678 & 0.848 & 0.955 & 0.880 &--  & 0.952 \\
\cline{2-9}
& 0.8 & 0.05 & 0.941 & 0.931 & 0.960 & 0.926 &--  & 0.937 \\
&     & 0.10 & 0.896 & 0.936 & 0.948 & 0.928 &--  & 0.956 \\
&     & 0.15 & 0.854 & 0.904 & 0.944 & 0.890 &--  & 0.942 \\
&     & 0.20 & 0.716 & 0.841 & 0.957 & 0.887 &--  & 0.955 \\
\cline{2-9}
& 0.9 & 0.05 & 0.934 & 0.937 & 0.936 & 0.930 &--  & 0.941 \\
&     & 0.10 & 0.907 & 0.940 & 0.942 & 0.925 &--  & 0.952 \\
&     & 0.15 & 0.853 & 0.915 & 0.936 & 0.886 &--  & 0.945 \\
&     & 0.20 & 0.799 & 0.874 & 0.958 & 0.900 &--  & 0.964 \\
\hline
\end{tabular}
\end{table}

\if0
Overdispersion coverage
\fi

\begin{table}[htbp]
\centering
\caption{
Comparisons of coverage probabilities between methods under model misspecification($N\in\{150,300\}$). Outcome data were generated from a Poisson distribution with over-dispersion, while inference was conducted dismissing it. The Bayesian joint approach shows notable under-coverage, whereas the proposed two-stage method with robust variance estimation achieves coverage close to or above the nominal 95\% level, at the cost of some conservativeness.
}

\begin{tabular}{c c c c c c c}
\hline
$N$ & \multicolumn{2}{c}{TRUE} & \multicolumn{2}{c}{BAYES} &  \multicolumn{2}{c}{PROPOSED} \\
\cline{2-7}
& $p_{\text{intact}}$ & $p_{\text{low}}$ 
& $p_{\text{intact}}$ & $p_{\text{low}}$ 
& $p_{\text{intact}}$ & $p_{\text{low}}$ \\
\hline
\multirow{12}{*}{150}
& 0.7 & 0.05 & 0.869 & 0.920 & 0.984 & 0.972\\
&     & 0.10 & 0.832 & 0.887 & 0.975 & 0.967\\
&     & 0.15 & 0.761 & 0.864 & 0.979 & 0.984\\
&     & 0.20 & 0.693 & 0.811 & 0.971 & 0.985\\
\cline{2-7}
& 0.8 & 0.05 & 0.838 & 0.916 & 0.967 & 0.960\\
&     & 0.10 & 0.831 & 0.899 & 0.972 & 0.979\\
&     & 0.15 & 0.757 & 0.845 & 0.969 & 0.977\\
&     & 0.20 & 0.697 & 0.798 & 0.970 & 0.979\\
\cline{2-7}
& 0.9 & 0.05 & 0.859 & 0.937 & 0.914 & 0.966\\
&     & 0.10 & 0.853 & 0.917 & 0.922 & 0.966\\
&     & 0.15 & 0.844 & 0.909 & 0.927 & 0.970\\
&     & 0.20 & 0.803 & 0.886 & 0.937 & 0.975\\
\hline
\multirow{12}{*}{300}
& 0.7 & 0.05 & 0.845 & 0.925 & 0.977 & 0.978\\
&     & 0.10 & 0.808 & 0.853 & 0.987 & 0.967\\
&     & 0.15 & 0.718 & 0.833 & 0.980 & 0.972\\
&     & 0.20 & 0.579 & 0.739 & 0.980 & 0.987\\
\cline{2-7}
& 0.8 & 0.05 & 0.827 & 0.910 & 0.982 & 0.965\\
&     & 0.10 & 0.825 & 0.883 & 0.983 & 0.980\\
&     & 0.15 & 0.719 & 0.816 & 0.977 & 0.980\\
&     & 0.20 & 0.595 & 0.733 & 0.970 & 0.982\\
\cline{2-7}
&0.9 & 0.05 & 0.836 & 0.930 & 0.963 & 0.977\\
&    & 0.10 & 0.803 & 0.894 & 0.960 & 0.976\\
&    & 0.15 & 0.764 & 0.873 & 0.953 & 0.978\\
&    & 0.20 & 0.635 & 0.778 & 0.960 & 0.979\\
\hline
\end{tabular}
\end{table}

\if0
NegBin coverage
\fi

\begin{table}[htbp]
\centering
\caption{
Comparisons of coverage probabilities between methods under model misspecification($N\in\{150,300\}$). Outcome data were generated from a negative binomial distribution, but inference was conducted assuming a Poisson model. The Bayesian joint approach shows notable under-coverage, whereas the proposed two-stage method with robust variance estimation achieves coverage close to or above the nominal 95\% level, at the cost of some conservativeness.
}
\begin{tabular}{c c c c c c c}
\hline
$N$ & \multicolumn{2}{c}{TRUE} & \multicolumn{2}{c}{BAYES} &  \multicolumn{2}{c}{PROPOSED} \\
\cline{2-7}
& $p_{\text{intact}}$ & $p_{\text{low}}$ 
& $p_{\text{intact}}$ & $p_{\text{low}}$ 
& $p_{\text{intact}}$ & $p_{\text{low}}$ \\
\hline
\multirow{12}{*}{150}
& 0.7 & 0.05 & 0.718 & 0.871 & 0.969 & 0.976\\
&     & 0.10 & 0.683 & 0.798 & 0.975 & 0.976\\
&     & 0.15 & 0.615 & 0.733 & 0.977 & 0.980\\
&     & 0.20 & 0.529 & 0.642 & 0.956 & 0.967\\
\cline{2-7}
& 0.8 & 0.05 & 0.723 & 0.881 & 0.959 & 0.972\\
&     & 0.10 & 0.662 & 0.797 & 0.958 & 0.967\\
&     & 0.15 & 0.608 & 0.760 & 0.953 & 0.976\\
&     & 0.20 & 0.531 & 0.663 & 0.970 & 0.979\\
\cline{2-7}
& 0.9 & 0.05 & 0.764 & 0.911 & 0.859 & 0.965\\
&     & 0.10 & 0.759 & 0.887 & 0.876 & 0.970\\
&     & 0.15 & 0.701 & 0.867 & 0.874 & 0.963\\
&     & 0.20 & 0.653 & 0.807 & 0.862 & 0.961\\
\hline
\multirow{12}{*}{300}
& 0.7 & 0.05 & 0.706 & 0.861 & 0.979 & 0.962\\
&     & 0.10 & 0.663 & 0.798 & 0.970 & 0.977\\
&     & 0.15 & 0.548 & 0.678 & 0.979 & 0.965\\
&     & 0.20 & 0.408 & 0.568 & 0.983 & 0.987\\
\cline{2-7}
& 0.8 & 0.05 & 0.660 & 0.845 & 0.972 & 0.972\\
&     & 0.10 & 0.646 & 0.785 & 0.984 & 0.979\\
&     & 0.15 & 0.580 & 0.696 & 0.974 & 0.975\\
&     & 0.20 & 0.432 & 0.562 & 0.977 & 0.985\\
\cline{2-7}
 & 0.9 & 0.05 & 0.670 & 0.887 & 0.919 & 0.970\\
 &  & 0.10 & 0.676 & 0.855 & 0.917 & 0.966\\
 &  & 0.15 & 0.591 & 0.773 & 0.925 & 0.963\\
 &  & 0.20 & 0.499 & 0.676 & 0.932 & 0.967\\\hline
\end{tabular}
\end{table}

\section{Application to Real Data}
We tested the proposed method to data on Pallas’s squirrels \textit{Callosciurus erythraeus} collected in the Kanto region of Japan. \textit{C. erythraeus} is an arboreal mammal species originally distributed in Southeast Asia, particularly in Taiwan. This species has been introduced worldwide and poses epidemiological risks by acting as a vector for zoonotic pathogens, as well as causing conflicts with native species. \citep{Katahira2022} collected host individuals in Kanto to assess the current infection status of \textit{C. erythraeus} and found that the abundance of ectoparasite species was higher than previously reported.

We conducted two parasite survey protocols, combing and shaving, on captured squirrels and recorded the number of parasite individuals belonging to six species: \textit{Haemaphysalis flava}, \textit{Leptotrombidium}  spp., \textit{Enderleinellus kumadai},  
\textit{Neohaematopinus callosciuri}, \textit{Ceratophyllus anisus}, and \textit{ Ceratophyllus indages indages}. We then estimated the probability of sample destruction associated with the combing protocol ($p_{\text{intact}}$) together with the sample detection probability ($p_{\text{low}}$). Covariates for the abundance model included sex (male = 1, female = 0), total length (TL), head-body length (HBL), and parasite species represented by dummy variables. The sample sizes were $N_{\mathrm{ctr}}=60$ and $N_{\mathrm{trt}}=120$, respectively. All captured squirrels were randomly assigned to the two survey groups. Continuous covariates were standardized to have mean 0 and standard deviation 1 to improve numerical stability and facilitate interpretation.

We applied the proposed two-stage estimator and a hierarchical Bayesian joint model, assuming that $p_{\text{intact}}$ and $p_{\text{low}}$ were homogeneous across parasite species. Although this assumption may not strictly hold, it was adopted because of the limited sample size. Even under this assumption, the estimated average value of $p_{\text{low}}$ can be used to correct abundance estimates obtained from the low-cost combing protocol. For the proposed two-stage estimator, the log-likelihood functions were implemented in R and optimized using the BFGS algorithm implemented in the \texttt{optim} function. For the hierarchical Bayesian model, 40,000 MCMC iterations were performed after discarding the first 10,000 iterations as burn-in. All analyses were conducted in \texttt{R} (version 4.3) using the package \texttt{R2jags}. Trace plots were inspected to confirm MCMC convergence.

Table 3 summarizes the point estimates of each parameter obtained by two different methods, and their $95\%$ interval estimates. The point estimates indicate indicate that the abundance of \textit{E. kumadai} and \textit{N. callosciuri} are larger than the other four species, after adjusting for host covariates, coinciding with the previous report \citep{Katahira2022}. Further, results of 1-st stage estimator supports our original motivation that the calibration of the sample loss caused by the combing method is required before estimating the ${\ p}_\text{low}$. Although squirrel individuals collected from the fields are assigned to control and treatment group at random, estimated ${\ p}_\text{intact}$ is about 0.7, substantially lower than 1.0. It indicates that the number of detected parasite species by shaving method is reduces by approximately 30$\%$ after one-combing session, even when the relevant covariates (i.e. sex, parasite species, and the body conditions) are controlled for.  
 
Comparing results between the proposed method and the Bayesian joint model, the estimated value of ${\ p}_\text{intact}$ is slightly larger under the hierarchical Bayesian model compared to the proposed method, while the estimates of ${\ p}_\text{low}$ are nearly identical. This result is consistent with the simulation study: when $N$ and ${\ p}_\text{low}$ are small, the bias of the hierarchical Bayesian model is not severe, although the proposed method is more accurate. In contrast, the $95 \%$ confidence interval for ${\ p}_\text{low}$ obtained by the proposed method is about 50 $\%$ wider that the $95\%$ credible interval from the hierarchical Bayesian model. Together with the simulation results, the proposed method is expected to provides a more robust and reliable assessment of uncertainty, whereas the hierarchical Bayesian model tends to underestimate it.

\begin{table}[htbp]
\centering
\caption{Point estimates and interval estimates of parameters obtained by different methods.a) Point estimates of the parameters. Maximum likelihood estimates (MLE) and posterior means from the Bayesian hierarchical joint model (Post. Mean) are reported together with their standard errors (SE) and posterior standard deviations (Post. SD). For the two-stage estimator, two types of standard errors are presented: the usual standard error (Naïve SE) and the corrected standard error, we proposed (Robust SE). b) $95 \% $ Interval of parameters estimated by different methods. Joint Bayes $95 \% $ CI denotes the posterior credible interval obtained from the Bayesian joint hierarchical model. Naïve $95 \% $ denotes the conventional interval ignoring the first-stage uncertainty, and Proposed $95 \% $ denotes the robust interval that accounts for the first-stage estimation error.}
\begin{subtable}{\linewidth}
\centering
\caption{Point estimates}
\begin{tabular}{lccccc}
\toprule
 & Post.\ Mean & Proposed &
 Post.\ SD & Naïve SE & Robust SE\\
\midrule
(Intercept)& -2.057 & -1.974 & 0.369 & 0.339 & -- \\
Sex&  1.492 &  1.509 & 0.069 & 0.070 & -- \\
sp.E\_kumadai&  3.841 &  3.784 & 0.366 & 0.337 & -- \\
sp.M\_anisas&  0.846 &  0.747 & 0.430 & 0.405 & -- \\
sp.M\_indages&  1.746 &  1.609 & 0.392 & 0.365 & -- \\
sp.N\_callosciuri  &  4.568 &  4.489 & 0.364 & 0.335 & -- \\
sp.T\_gen& -1.732 & -1.504 & 0.883 & 0.782 & -- \\
TL&  0.072 &  0.079 & 0.029 & 0.029 & -- \\
HBL& -0.551 & -0.559 & 0.029 & 0.029 & -- \\
$p_{\mathrm{intact}}$& 0.715 & 0.700 & 0.028 & 0.131 & -- \\
$p_{\mathrm{low}}$& 0.040 & 0.041 & 0.005 & 0.120 & 0.196 \\
\bottomrule
\end{tabular}
\end{subtable}

\begin{subtable}{\linewidth}
\centering
\caption{95\% intervals}
\begin{tabular}{lccc}
\toprule
 & Bayesian & Naïve & Proposed\\
\midrule
(Intercept)& $-2.894,\,-1.402$ & $-2.639,\,-1.308$ & -- \\
Sex& $\phantom{-}1.357,\,\phantom{-}1.630$   & $\phantom{-}1.371,\,\phantom{-}1.647$   & -- \\
sp.E\_kumadai& $\phantom{-}3.199,\,\phantom{-}4.664$   & $\phantom{-}3.123,\,\phantom{-}4.445$   & -- \\
sp.M\_anisas& $\phantom{-}0.050,\,\phantom{-}1.760$   & $-0.046,\,\phantom{-}1.540$  & -- \\
sp.M\_indages& $\phantom{-}1.034,\,\phantom{-}2.604$   & $\phantom{-}0.894,\,\phantom{-}2.325$   & -- \\
sp.N\_callosciuri & $\phantom{-}3.927,\,\phantom{-}5.375$   & $\phantom{-}3.832,\,\phantom{-}5.146$   & -- \\
sp.T\_gen& $-3.699,\,-0.220$ & $-3.036,\,\phantom{-}0.028$  & -- \\
TL& $\phantom{-}0.016,\,\phantom{-}0.130$   & $\phantom{-}0.021,\,\phantom{-}0.136$   & -- \\
HBL& $-0.609,\,-0.494$ & $-0.616,\,-0.502$ & -- \\
$p_{\mathrm{intact}}$& $\phantom{-}0.663,\,\phantom{-}0.771$   & $\phantom{-}0.643,\,\phantom{-}0.751$   & -- \\
$p_{\mathrm{low}}$& $\phantom{-}0.031,\,\phantom{-}0.050$& $\phantom{-}0.033,\,\phantom{-}0.051$   & $\phantom{-}0.028,\,\phantom{-}0.059$ \\
\bottomrule
\end{tabular}
\end{subtable}
\end{table}

\section{Discussion}
In this article, we developed a two-stage estimator for abundance that is applicable in situations where two distinct survey protocols interact in a way that alters the detection probability of one of the methods. We further derived a robust, hamburger-type variance estimator that extends the conventional sandwich variance to account for the propagation of uncertainty from the first-stage estimation. Applying our method to empirical parasite data, we demonstrated that combining a high-accuracy shaving procedure with a low-cost combing method leads to an overestimation of the detection probability of the latter.

A key advantage of our method is that it separates the estimation procedure into two different stages, by which heterogeneity of the detection probability can be corrected. Our proposed method showed good performances in the simulation experiment, both in point estimation and in the assessment of uncertainty. In contrast, the Bayesian hierarchical joint model did not perform well in our numerical experiments, in terms of both point estimation and uncertainty quantification, even when the model was correctly specified. Specifically, the joint model yielded biased estimates of $p_{\text{intact}}$ and $p_{\text{low}}$, and the coverage probabilities of the nominal $95\%$ credible intervals were substantially below the nominal level when $N = 150$. Increasing the sample size to $N = 300$ did not lead to a meaningful improvement. These results indicate that, under the data-generating mechanism considered here, the joint model may be less reliable than the proposed two-stage approach. The theoretical reasons for this behavior are not fully understood and are beyond the scope of the present study; in particular, alternative MCMC implementations, different prior specifications, or other joint model formulations might lead to improved performance.

Several limitations remain to be addressed in future work. First, although we have focused on an i.i.d.\ setting, autocorrelation structures may pose difficulties for broader applications. Our empirical example is based on a randomized controlled design, which ensures independence across subjects; however, spatio-temporal dependence is common in ecological abundance data yielding to misleading inferences if ignored. Developing procedures that explicitly accommodate such dependence—potentially through extensions based on generalized method-of-moments (GMM) theory or correlated estimating equations—would considerably enhance the applicability of the proposed estimator.

Second, while extending the detection model to incorporate covariates is straightforward within our framework—because the proposed robust variance directly applies to a logistic-regression formulation—further generalization may still be beneficial. In particular, real-world surveys often exhibit heterogeneity arising from behavioral, physiological, or environmental factors that cannot be fully captured by fixed covariates alone. Allowing detection probabilities to vary through random effects, hierarchical structures, or latent classes may provide additional flexibility and improve robustness against model misspecification, especially in systems where detection processes are complex or partially unobserved.

Although several issues remain for future improves, the present work establishes a flexible foundation that can be readily extended to a wide range of ecological monitoring programs. We hope that this framework contributes not only to improving the accuracy of abundance estimation in mixed-survey designs but also to encouraging more rigorous treatment of uncertainty in ecological inference.

\clearpage

\section*{Appendix}
In this appendix, we formally derive a robust variance estimator for the second‑stage parameter $\widehat{p}_{\mathrm{low}}$. The key insight is that the estimation error in the first stage propagates into the second‑stage estimating equation. Our derivation follows the general asymptotic theory of M-estimators \citep{Huber1967,Huber1973,NeweyMcFadden1994} and the two-step estimation framework of \citep{MurphyTopel2002}, who showed that ignoring first-stage uncertainty leads to systematically underestimated standard errors. By applying a first‑order Taylor expansion around the true parameter values, we explicitly characterize how this additional variability enters the asymptotic variance of $\widehat{p}_{\mathrm{low}}$. This derivation extends the conventional ``sandwich'' robust variance estimator (\citep{Eicker1963,White1980}) to incorporate first‑stage uncertainty, yielding what we refer to as the ``hamburger'' robust variance. It corrects the naive standard error formula that would otherwise ignore this additional contribution.

\vspace{0.5em}
\subsection*{Setup and notation}

In the second stage of our estimator, the key parameter of interest is the 
detection probability of the low–accuracy method (e.g.\ the combing survey).  
Let \(p\) denote the per–session detection probability, and introduce its 
logit–scale parameter
\begin{align}
p = \operatorname{logit}^{-1}(\eta).
\end{align}

For an individual with effort level \(Z_i \in \{0,1,2,\dots\}\), the cumulative detection probability after \(Z_i\) repeated sessions is
\begin{align}
q_X(p) = 1 - (1-p)^{Z_i},
\end{align}
which corresponds to the probability that the individual is detected in at least one of the \(Z_i\) independent trials. The expected count contributed by individual \(i\) in the Poisson model is then
\begin{align}
\mu_i(\eta,\beta)
= \lambda_i(\beta)\, q_X(p),
\end{align}
where, \(\lambda_i(\beta)\) is the predicted true abundance obtained from the first-stage model, which is the expected number of individuals detected by the low–accuracy method, conditional on the estimate from the first stage. The individual log-likelihood contribution under a Poisson model is
\begin{align}
\ell_i(\eta,\beta)
  = Y_i^{\text{low}}\log \mu_i(\eta,\beta)
    - \mu_i(\eta,\beta)
    - \log(Y_i^{\text{low}}!).
\end{align}

Let denote the individual score function and its average score as:
\begin{align}
u_i(\eta,\beta)
  = \frac{\partial}{\partial\eta}\,\ell_i(\eta,\beta).
\end{align}

\begin{align}
U_n(\eta,\beta)=\frac{1}{n}\sum_{i=1}^n u_i(\eta,\beta).
\end{align}

Denote the true parameter values by \((\eta_0,\beta_0)\), and let \(\hat\eta\) solve the
estimating equation
\begin{align}
U_n(\hat\eta,\hat\beta)=0.
\end{align}

For notational convenience, define the derivative matrices evaluated at the true values
\begin{align}
A_n\equiv \left.\frac{\partial U_n(\eta,\beta)}{\partial\eta^\top}\right|_{(\eta_0,\beta_0)},\qquad
C_n\equiv \left.\frac{\partial U_n(\eta,\beta)}{\partial\beta^\top}\right|_{(\eta_0,\beta_0)}.
\end{align}
In addition, we follow the usual sandwich variance notation and define the empirical "meat" matrix of the second-stage scores
(on the average-score scale):
\begin{align}
B_n \equiv \frac{1}{n}\sum_{i=1}^n u_i(\eta_0,\beta_0)\,u_i(\eta_0,\beta_0)^\top.
\end{align}

\subsection*{Assumptions}
We impose the following regularity conditions:

\begin{enumerate}
  \item[(A1)] The first–stage estimator $\hat\beta$ is obtained as the solution of a regular estimating equation satisfying the usual regularity conditions for M-estimation.

  \item[(A2)] For each $i$, the map 
  \[
     (\eta,\beta)\mapsto u_i(\eta,\beta)
  \]
  is twice continuously differentiable in a neighborhood of $(\eta_0,\beta_0)$.

  \item[(A3)]  There exist finite matrices \(A, C\) such that
\[
A_n \to_p A,\qquad C_n \to_p C,
\]
where \(A\) is nonsingular.

  \item[(A4)] The second derivatives of $U_n(\eta,\beta)$ are uniformly bounded 
  in a neighborhood of $(\eta_0,\beta_0)$.
  \item[(A5)] The sequence $\{u_i(\eta_0,\beta_0)\}$ satisfies a central limit theorem:
\[
   \frac{1}{\sqrt{n}}\sum_{i=1}^n u_i(\eta_0,\beta_0)
      \;\overset{d}{\longrightarrow}\; N(0,B),
\]
where the asymptotic covariance matrix $B$ is given by
\[
   B = \operatorname{plim}_{n\to\infty} B_n.
\]
\end{enumerate}
\vspace{0.5em}

We perform a first–order Taylor expansion of \(U_n(\hat\eta,\hat\beta)\) around 
\((\eta_0,\beta_0)\):
\begin{align}
0 = U_n(\hat\eta,\hat\beta)
   = U_n(\eta_0,\beta_0)
     + A_n(\hat\eta-\eta_0)
     + C_n(\hat\beta-\beta_0)
     + R_n,
\end{align}
where \(R_n\) collects the higher–order terms, giving a rearranged form as
\begin{align}
\sqrt n\,(\hat\eta-\eta_0)
= -A_n^{-1}\Bigl\{
     \sqrt n\,U_n(\eta_0,\beta_0)
     + C_n\sqrt n\,(\hat\beta-\beta_0)
     + \sqrt n\,R_n
   \Bigr\}.
\end{align}

Since
\begin{align}
\sqrt n\,U_n(\eta_0,\beta_0)
  = \frac{1}{\sqrt n}\sum_{i=1}^n u_i(\eta_0,\beta_0),
\end{align}
and using \(A_n^{-1}\overset{p}{\to}A^{-1}\), 
\(C_n\overset{p}{\to}C\), and \(\sqrt n\,R_n=o_p(1)\),  
we obtain the asymptotic linear representation
\begin{align}
\sqrt n\,(\hat\eta-\eta_0)
  = -A^{-1}\!\left\{
        \frac{1}{\sqrt n}\sum_{i=1}^n u_i(\eta_0,\beta_0)
        + C\,\sqrt n\,(\hat\beta-\beta_0)
     \right\}
     + o_p(1).
\end{align}

This decomposition shows two sources of asymptotic variation in \(\hat\eta\):
the direct stochastic component 
\(\frac{1}{\sqrt n}\sum_{i=1}^n u_i(\eta_0,\beta_0)\), which satisfies a central limit theorem, and the additional variability induced by estimation of the first–stage
parameter \(\beta\), captured through the term 
\(C\,\sqrt n(\hat\beta-\beta_0)\).

Since \(\hat\beta\) is itself obtained from a regular first–stage estimating
equation, assuming the first–stage sample size \(N\) satisfies \(n/N \to c \in (0,\infty)\), it admits an asymptotic linear representation. 
\[
\sqrt n\,(\hat\beta-\beta_0)
  = \frac{1}{\sqrt n}\sum_{i=1}^n \psi_i + o_p(1),
\]
for some influence functions \(\psi_i\), which is scaled so that the expansion is expressed in the \(n\)-based normalization appropriate for the second stage.

Substituting this expression into the expansion for the second stage yields
\begin{align}
\sqrt n\,(\hat\eta-\eta_0)
  = -A^{-1}\frac{1}{\sqrt n}
      \sum_{i=1}^n \bigl\{
         u_i(\eta_0,\beta_0)
         + C\,\psi_i
      \bigr\}
    + o_p(1).
\end{align}

Therefore,
\begin{align}
\sqrt n\,(\hat\eta-\eta_0)
  \;\overset{d}{\longrightarrow}\;
  N\!\left(0,\;V_\eta\right),
\end{align}
where the asymptotic variance is
\begin{align}
V_\eta
  = A^{-1}\bigl(
      B + C\,V_\beta\,C^\top
    \bigr)A^{-\top},
\end{align}
with 
\(B = \operatorname{Var}\{u_i(\eta_0,\beta_0)\}\) and
\(V_\beta=\operatorname{Var}(\psi_i)\). This robust ``hamburger'' variance estimator thus provides a correction to standard errors in two‑stage estimation when the first‑stage uncertainty is non‑negligible. If \(\beta\) were known (i.e.\ \(\hat\beta=\beta_0\)), the additional term involving \(C\) would vanish, and the variance reduces to the usual sandwich form \(A^{-1} B A^{-\top}\).

\end{document}